# Phase Transitions in a Dusty Plasma with Two Distinct Particle Sizes

B. Smith, T. Hyde, L. Matthews, J. Reay, M. Cook, and J. Schmoke

Center for Astrophysics, Space Physics and Engineering Research (CASPER), Baylor University, One Bear Place #97310, Waco, TX 76798-7310, USA
Email:  Truell_Hyde@baylor.edu

## Abstract

In semiconductor manufacturing, contamination due to particulates significantly decreases the yield and quality of device fabrication, therefore increasing the cost of production.  Dust particle clouds can be found in almost all plasma processing environments including both plasma etching devices and in plasma deposition processes.  Dust particles suspended within such plasmas will acquire an electric charge from collisions with free electrons in the plasma.  If the ratio of inter-particle potential energy to the average kinetic energy is sufficient, the particles will form either a "liquid" structure with short range ordering or a crystalline structure with long range ordering.  Otherwise, the dust particle system will remain in a gaseous state.  Many experiments have been conducted over the past decade on such colloidal plasmas to discover the character of the systems formed, but more work is needed to fully understand these structures.  The preponderance of previous experiments used monodisperse spheres to form complex plasma systems.  However, most plasma processing environments contain more arbitrary distributions of particle size.  In order to examine in more detail the effects of a size distribution, experiments were carried out in a GEC rf reference cell modified for use as a dusty plasma system.  Using two monodisperse particle sizes, experiments were conducted to determine the manner in which phase transitions and other thermodynamic properties depended upon the overall dust grain size distribution.  Two dimensional (2D) plasma crystals were formed from different mixtures of 8.89 µm and 6.50 µm monodisperse particles in Argon plasma.  With the use of various optical techniques, the pair correlation function was determined at different pressures and powers and then compared to measurements obtained for monodisperse spheres.

## 1. Introduction

Nearly all environments which contain plasma also contain dust.  Dust can be found in plasma in vastly different environments ranging from protostellar clouds to plasma processing environments.  Dusty plasma systems can also be found in planetary rings, cometary environments, and the Earth's ionosphere and magnetosphere.  As a result, the emerging field of dusty or colloidal plasmas has witnessed a dramatic rise in interest over the past decade.  Under certain conditions these dust particles immersed in plasma environments can form Coulomb crystals.  Such man-made plasma crystal systems model some of the naturally occurring systems above and mirror the structure, dynamics, and phase transitions of condensed matter physics.

Through the processes of ion and electron collisions, photoemission, and secondary electron emission, dust immersed in plasma acquires an electric charge.  Secondary electron emission and photoemission can be ignored in low temperature plasma, especially if ultraviolet light is absent. In such an environment, grain charging depends only on the ion and primary electron currents. The dust grain charge is normally negative, since the incident flux of the faster electrons from the enveloping plasma is higher (Morfill et al., 1999).

A dusty or colloidal plasma can be loosely defined as a plasma with a dust component, where the dust can no longer be considered a contaminant but must instead be considered an integral part of the plasma. Coulomb crystals require the dusty or colloidal plasma to reach a state where the dust crystals begin to show short and sometimes even long range ordering. Once the dust particle number density is high enough to meet the above criteria, the dust can form these crystalline structures. The Coulomb coupling parameter ($\Gamma$) determines how "solid" these colloidal plasmas become. $\Gamma$ is the ratio of the interparticle Coulomb potential energy to the thermal energy of the particles. If $\Gamma$ becomes greater than some critical value ($\Gamma_{Crit}$), the system behaves like a solid (Ichimaru, 1982). (In other words, long-range ordering is present.) For cases where $\Gamma$ is comparable to or less than $\Gamma_{Crit}$ short-range ordering dominates and the crystal is said to be in a liquid phase (Thomas et al., 1994).

Many experimenters have used monodisperse spheres in single component plasmas to examine Coulomb crystals. See for example Pieper et al. (1996a), Pieper et al. (1996b), and Zuzic et al. (2000). However, in this experiment, we seek to discover how a distribution of particle sizes might affect the crystalline structure of a colloidal plasma.

## 2. Experiment

These experiments were conducted in a Gaseous Electronics Conference (GEC) radio frequency (rf) Reference Cell (Hargis et el., 1994). The upper electrode of the GEC rf Reference Cell was replaced with a grounded ring electrode with the walls of the cell serving as grounds. To allow the dust crystal to be imaged from above, an optical window was installed in the upper electrode port with a feed-through for dust particles. The lower electrode was capacitively coupled to an rf frequency generator and amplifier through a T-type matching network. The lower electrode has a stainless steel sheath ground shield and a Teflon insulator. An aluminum dish with a circular cutout rests on the lower electrode. The cutout establishes the boundary conditions, constraining the dust particles to remain above the electrode.

The dust particles were illuminated using a horizontal diode laser sheet, created by a LASIRIS™ 685 nm, 50 mW diode laser which had a LASIRIS™ 5° fan angle, line generating lens attached. This sheet was then imaged using a CCD (charge-coupled device) camera (Cohu™ 7800 series) which was vertically mounted above the cell. A 680 nm band pass filter (Melles Griot™ 03FIV325) was used on the camera to capture only the reflected laser light. This was necessary to block the light generated by the plasma itself. The camera was equipped with a zoom lens (Edmund Industrial Optics™ #52-274 Close Focus Zoom Lens) which allows for close inspection of the dust crystal. The number of particles contained in each image varied, but averaged around 1000 particles.

The purpose of this set of experiments was to study the differences between monodisperse particle distributions and two component particle distributions in dusty plasmas. The two monodisperse particle distributions chosen for this experiment were 8.89±0.09 µm and 6.50±0.08 µm. The dusty plasma was generated at 50 mTorr, 100 mTorr, 200 mTorr, 400 mTorr, 600 mTorr, 800 mTorr, and 1 Torr. For each pressure the applied voltage was varied from 45 Vpp to 80 Vpp, in steps of approximately 11 V, to see the effects (if any) on the crystal at that pressure. Sets of data were taken for a pure 8.9 μm distribution, a pure 6.5 μm distribution, and a 50/50 mix of 8.9 μm and 6.5 μm particles. The resulting images were analyzed using MATLAB, and a pair correlation function, g(r), was generated for each run. These graphs were then compared to one another and to graphs obtained theoretically.

## 3. Results

The data presented here is a small subset of the total data taken; however these results are representative of the whole. The results at 100 mTorr represent the data taken from 50 to 200 mTorr, the results at 400 mTorr represent the data taken from 200 mTorr to 800 mTorr, and the results at 1 Torr represent the data taken from 800 mTorr to 3 Torr. The powers shown here represent the lowest three successive powers used at each pressure.

The pair correlation function or radial distribution function, g(r), is useful in determining the amount and type of ordering present in a crystalline lattice. This function has been used for many years in condensed matter physics, but has also been used in two-dimensional and three-dimensional dusty plasma physics. (See for example the work by Quinn and colleagues (1996) and by Zuzic et al. (2000).) The pair correlation functions represented here have been normalized by the annular area as shown elsewhere (Smith et al., 2004). However, the positions have not been normalized to set the nearest neighbor distance to one. This allows one to examine the shift in nearest neighbor distance as pressures and powers are changed.

At 100 mTorr, none of the distributions showed good crystalline characteristics at any power, as shown in Figures 1-3. However, the 8.9 μm and the 6.5 μm distributions did transition from a more ordered liquid state (possibly a liquid crystal) to a less ordered one as the power was increased. This is demonstrated by the decrease in the number of peaks present in Figures 1 and 2. Both of these distributions also show a similar trend in the shift of the nearest neighbor peaks (first maximum in Figures 1 and 2) to smaller values as the power is increased. Alternatively, the 50/50 mix shows little transition or nearest neighbor shift as power is increased, as seen in Figure 3.

At 400 mTorr, the three distributions all show more ordering than they did at 100 mTorr. As can be seen in Figures 4 and 5, both the 8.9 μm and the 6.5 μm distributions start in a liquid crystal state and as the power is increased become more ordered. This is evidenced by the fact that the peaks in Figures 3 and 4 become more distinct and increase in number. The 50/50 mix becomes slightly more ordered as the power increases, but remains in a liquid state for all powers (Figure 6). The nearest neighbor shift is approximately the same for all three distributions at this power and equals about 18 μm total for each graph (Figures 4-6).

At 1 Torr, all three distributions show good ordering for all powers (Figures 7-9). However, the 8.9 μm and the 6.5 μm distributions (Figures 7 and 8) exhibit definite crystal structure at all powers, whereas the 50/50 mix (Figure 9) starts as a less ordered liquid crystal. There are distinct differences in the nearest neighbor shifts for each of the distributions as well. The 8.9 μm distribution has the largest shift at 35 μm (Figure 7), while the 50/50 mix has the least at 14 μm (Figure 9).

## 4. Conclusions

In this work, it has been shown that having a two component mix of dust distributions does affect the phase transitions within the particle cloud. While both of the monodisperse distributions exhibit Coulomb crystallization over a wide range of powers and pressures, the 50/50 mix requires higher pressure to begin to show crystallization. Additionally at 100 mTorr, mixing the 8.9 μm and 6.5 μm particles causes a definite phase change at 2 W. At 400 mTorr, the 50/50 mix seems to have characteristics of both the pure 8.9 μm and the pure 6.5 μm distributions. This suggests the 50/50 mix is undergoing some vertical separation due to the charge to mass ratio differences between the two particle sizes, and the differently sized particles are now only weakly correlated. Finally at 1 Torr, all of the distributions seem to have similar characteristics. This suggests the 6.5 μm and 8.9 μm particles may have completely separated into different layers; therefore, only one distribution can be viewed two-dimensionally.

In future, this group plans to examine the three-dimensional structure of dust crystals under the same set of conditions. Three-dimensional data will allow the separation into layers to be examined more carefully for both monodisperse distributions and for the various mixes.

## Acknowledgements

The authors would like to acknowledge the following people for their tireless work on this project: Benjamin Brooks, Aaron Jesseph, Shelly Johnson, Eugene Marinelli, Desireé Miles, Gary Shetler, and Jared Templeton.

## References

Hargis, P. J., K. E. Greenberg, P. A. Miller et al., The Gaseous Electronic Conference radio-frequency reference cell: a defined parallel-plate radio-frequency system for experimental and theoretical studies of plasma-producing discharges, *Rev. Sci. Instrum.*, **65**, 140-154, 1994.

Ichimaru, S., Strongly coupled plasmas: High density classical plasmas and degenerate electron liquids, *Rev. Mod. Phys.*, **54**, 1017-1059, 1982.

Morfill, G. E., H. M. Thomas, U. Konopka et al., The plasma condensation: liquid and crystalline plasmas, *Phys. Plasmas*, **6**, 1769-1780, 1999.

Pieper, J. B., J. Goree, and R. A. Quinn, Experimental studies of two-dimensional and three-dimensional structure in a crystallized dusty plasma, *J Vac. Sci. Technol. A*, **14**, 519-524, 1996a.

Pieper, J. B., J. Goree, and R. A. Quinn, Three-dimensional structure in a crystallized dusty plasma, *Phys. Rev. E*, **54**, 5636-5640, 1996b.

Quinn, R. A., C. Cui, J. Goree et al., Structural analysis of a Coulomb lattice in a dusty plasma, *Phys. Rev. E*, **53**, R2049-R2052, 1996.

Smith, B., J. Vasut, T. Hyde et al., Dusty plasma correlation experiment, Accepted for publication in *Adv. Space Res.*, 2004.

Thomas, H., G. E. Morfill, V. Demmel et al., Plasma crystal: Coulomb crystallization in a dusty plasma, *Phys. Rev. Lett.*, **73**, 652-655, 1994.

M. Zuzic, A. V. Ivlev, J. Goree et al., Three-dimensional strongly coupled plasma crystal under gravity conditions, *Phys. Rev. Lett.*, **85**, 4064-4067, 2000.

Figure Captions

Figure 1. Pair correlation function for 8.9 μm particles in 100 mTorr argon plasma at three different plasma powers.

Figure 2. Pair correlation function for 6.5 μm particles in 100 mTorr argon plasma at three different plasma powers.

Figure 3. Pair correlation function for 50/50 mix of 6.5 μm and 8.9 μm particles in 100 mTorr argon plasma at three different plasma powers.

Figure 4. Pair correlation function for 8.9 μm particles in 400 mTorr argon plasma at three different plasma powers.

Figure 5. Pair correlation function for 6.5 μm particles in 400 mTorr argon plasma at three different plasma powers.

Figure 6. Pair correlation function for 50/50 mix of 6.5 μm and 8.9 μm particles in 400 mTorr argon plasma at three different plasma powers.

Figure 7. Pair correlation function for 8.9 μm particles in 1 Torr argon plasma at three different plasma powers.

Figure 8. Pair correlation function for 6.5 μm particles in 100 mTorr argon plasma at three different plasma powers.

Figure 9. Pair correlation function for 50/50 mix of 6.5 μm and 8.9 μm particles in 100 mTorr argon plasma at three different plasma powers.

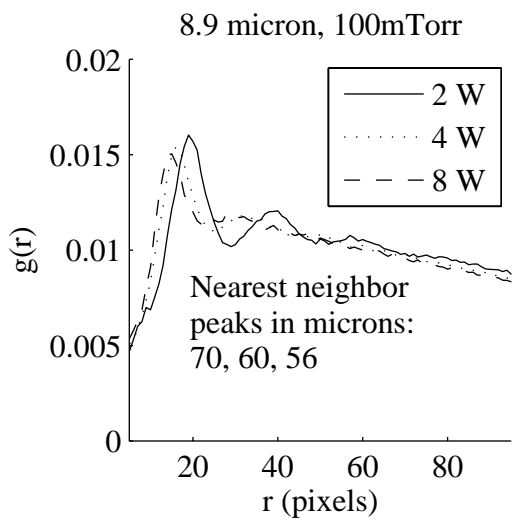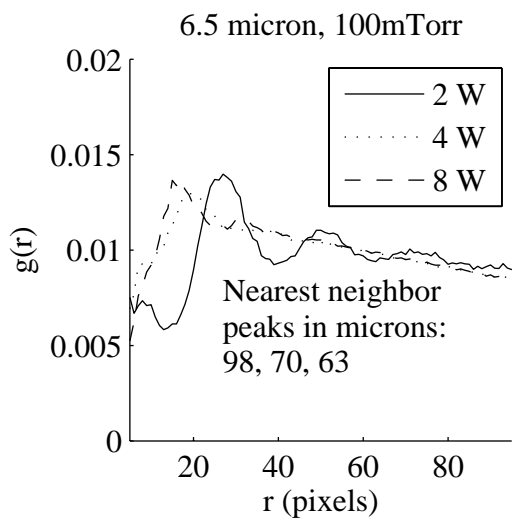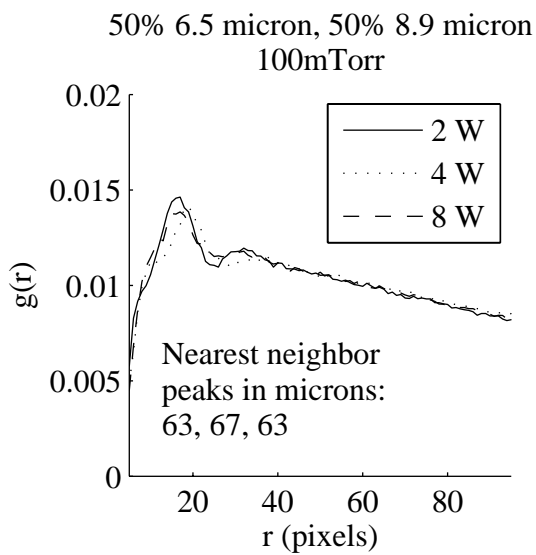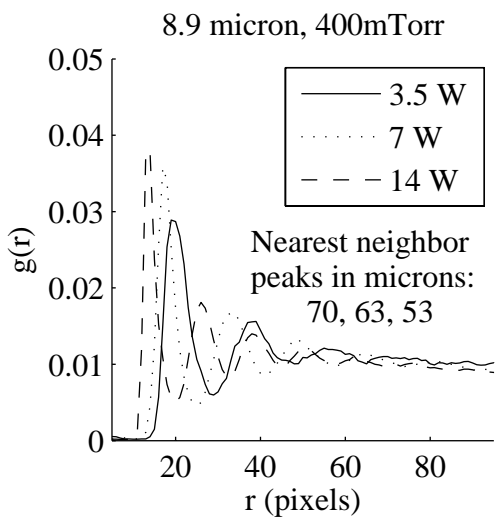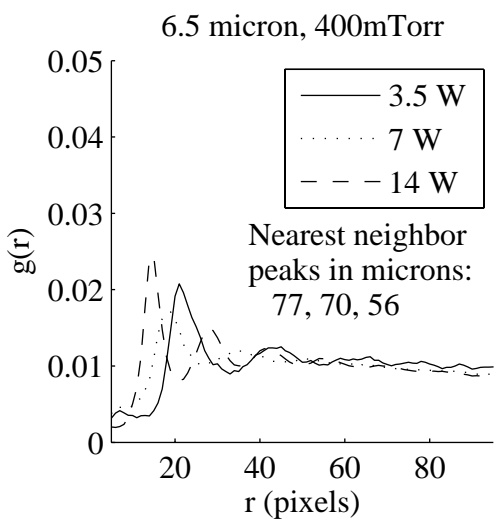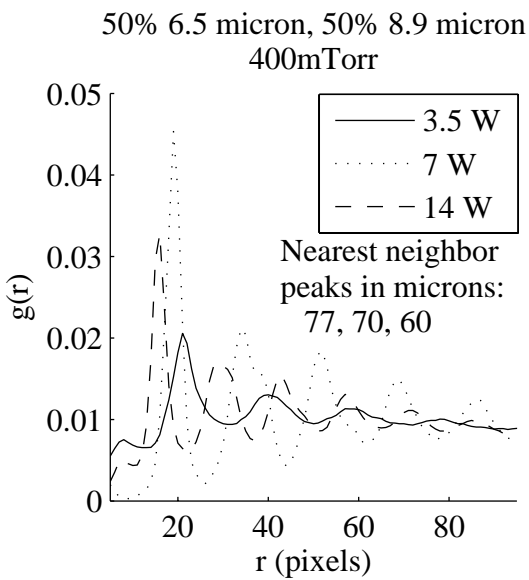

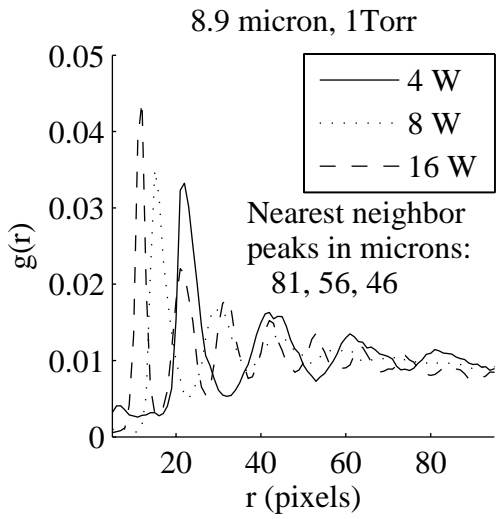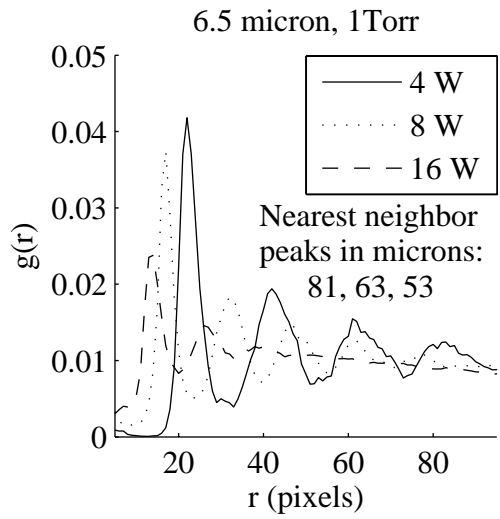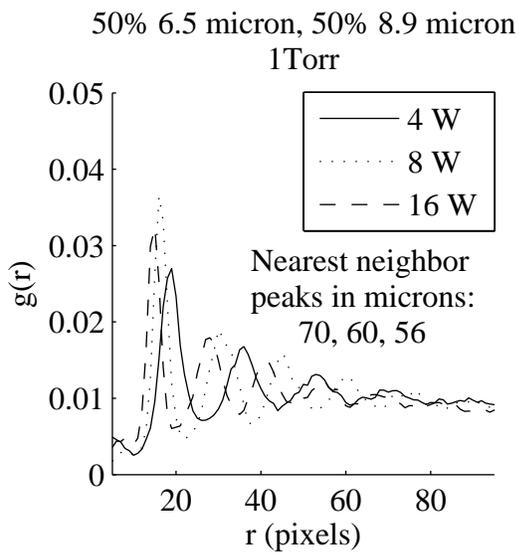